\documentclass[12pt]{article}

\usepackage{amssymb, amsmath}
\usepackage{graphicx}
\usepackage{cite}

\newcommand{\td}{\text{d}}

\def\be{\begin{equation}}
\def\ee{\end{equation}}
\def\bea{\begin{eqnarray}}
\def\eea{\end{eqnarray}}

\title{ \bf{The first law of soliton and black hole mechanics in five dimensions}}

\author{Hari K. Kunduri$^a$\footnote{hkkunduri@mun.ca } \  and James Lucietti$^b$\footnote{j.lucietti@ed.ac.uk } \\ \\
\small \sl $^a$ Department of Mathematics and Statistics, \\  \small \sl Memorial University of Newfoundland \\ \small \sl St John's NL A1C 4P5, Canada
\\ \small \sl $^b$  School of Mathematics and Maxwell Institute of Mathematical Sciences, \\ \small \sl University of Edinburgh, \\ \small \sl   King's Buildings, Edinburgh, EH9 3JZ, UK }

\date{}

\begin{document}

\maketitle


\begin{abstract}
We derive a mass formula and a mass variation law for asymptotically flat, stationary spacetimes, invariant under two commuting rotational symmetries, in a general five dimensional theory of gravity coupled to an arbitrary set of Maxwell fields and uncharged scalar fields. If the spacetime is everywhere regular, these mass formulas reduce to a sum of magnetic flux terms defined on its non-trivial 2-cycles. If there is a black hole, we obtain a mass variation law more general than previously obtained, which also has contributions from the 2-cycles exterior to the black hole. This can be interpreted as the first law of black hole mechanics in a background soliton containing bubbles.
\end{abstract}
\section{Introduction and main results}

Stationary spacetimes in general relativity are of central importance. The simplest finite energy examples contain black holes, which typically possess a singularity behind their horizon. It is natural to wonder whether everywhere regular stationary spacetimes with finite energy, termed {\it  solitons}, exist. In four dimensional Einstein-Maxwell type theories, asymptotically flat solitons are forbidden by the Lichnerowicz theorem, and generalisations thereof, under the assumption of strict stationarity (i.e. the stationary Killing field is timelike everywhere). For this reason it was often thought that extreme black holes are the closest gravitational analogue of a soliton, summarised by the slogan ``no solitons without horizons"~\cite{Gibbons:1990um, Gibbons:1997cc}.

In fact the Lichnerowicz theorem is closely related to certain intermediate results used to prove the black hole uniqueness theorem. Hence it typically fails in theories for which black hole uniqueness does. The first example of this was in four dimensional Einstein-Yang-Mills theory, which remarkably admits  asymptotically flat solitons as well as  associated ``hairy" black holes~\cite{Volkov:1998cc}. The mechanism responsible for their existence appears to originate from non-trivial topology of the gauge group~\cite{Sudarsky:1992ty}, which allows for self-sourcing finite energy field configurations with no need for a singularity.

In higher dimensions, there exists another mechanism by which solitons can be supported: non-trivial {\it spacetime topology}. For example, solitons in Kaluza-Klein theory are easy to construct, the simplest instance being the static KK bubble (i.e. $\mathbb{R}_t\times$ Euclidean Schwarzschild). In this note we consider {\it asymptotically flat} spacetimes.  Topological censorship implies that the domain of outer communication of any globally hyperbolic asymptotically flat spacetime satisfying a null energy condition must be simply connected~\cite{Topcen}. It follows that any spatial hypersurface $\Sigma$ must also have trivial fundamental group. In four spacetime dimensions this is sufficient to imply that the three-manifold $\Sigma$ has trivial homology groups. However, in higher dimensions, simple connectedness is less of a constraint. In particular, one may have non-trivial higher homology groups $H_p(\Sigma)$ for $p\geq 2$.

In fact,  the Lichnerowicz theorem for asymptotically flat vacuum spacetimes easily generalises to higher dimensions, although it is worth noting that this does not rule out the possibility of solitons for which the stationary Killing field becomes null or even spacelike at least somewhere in the interior. Furthermore, one could envisage that a $p$-form field could indeed ``support" non-trivial $p$-cycles with a magnetic flux.  In fact for Einstein-Maxwell theory, or Einstein gravity coupled to higher rank $p$-forms, a Lichnerowicz type theorem has been proved~\cite{Shiromizu:2012hb}, although inspection of their proof reveals that it is only valid if one assumes  trivial spacetime topology (they assume global existence of certain potentials).

Strikingly, explicit examples of  asymptotically flat soliton spacetimes with non-trivial 2-cycles, often termed ``bubbles", are known to exist in five dimensional supergravity theories~\cite{Giusto:2004kj, Bena:2005va, Berglund:2005vb}. These examples are supersymmetric and by now large families of such BPS-solitons have been constructed by exploiting a hidden linearity of the BPS-equations.  Remarkably, nonsupersymmetric solitons in such five dimensional theories have also been constructed~\cite{Bena:2009qv, Compere:2009iy, Bobev:2009kn}.  These solitons have been of particular interest due to the `fuzzball' conjecture (see e.g. \cite{Mathur, Mathur2}) in which they are interpreted as semi-classical microstates of black holes.  Furthermore, one can also construct asymptotically flat ``hairy" black hole solutions representing the superposition of a black hole and external bubbles, which may be intepreted as a black hole sitting in a soliton background, see e.g.~\cite{Bena:2011zw}.

Recently, Gibbons and Warner~\cite{Gibbons:2013tqa} derived a mass formula (Smarr relation) for such solitons in five dimensional $U(1)^N$-supergravity, which reveals why the Lichnerowicz theorem fails. Usually, certain potentials are introduced to derive the Smarr relation, which only exist if $H_2(\Sigma)=0$. By cleverly avoiding using such gauge potentials they were able to deal with a general simply connected spacetime, revealing the existence of new terms which represent the contribution of the various bubbles to the total mass. Hence the absence of horizons does not imply zero mass, so the positive mass theorem cannot be invoked to deduce the spacetime is flat. \footnote{Note that the known solitons are not {\it strictly} stationary. They possess time-like hypersurfaces in which the stationary Killing field may become null, so this is another way they evade the Lichnerowicz theorems.}

The original derivation of the Smarr relation and first law of black hole mechanics in five dimensional Einstein-Maxwell theories assumed the existence of a globally defined gauge potential~\cite{Gauntlett:1998fz}, hence it is only valid in general for trivial spacetime topology. With the discovery of black rings it became clear that, on top of the usual terms corresponding to the conserved charges of the spacetime, an extra term corresponding to the magnetic flux through an $S^2$ that links the ring, termed a ``dipole charge", must appear~\cite{Emparan:2004wy}. The presence of this extra term has been identified in a more general derivation of the first law by relaxing the assumption that the gauge potential is regular on the horizon~\cite{Copsey:2005se}. In fact, derivations of the first law in higher dimensional theories with $p$-forms typically reveal terms arising from non-trivial cycles on the horizon~\cite{Compere:2007vx, Hollands:2012cc, Keir:2013jga}. However, even these derivations do not allow for a general spacetime topology, and given the recent results of~\cite{Gibbons:2013tqa}, one may expect contributions to the first law from bubbles outside the black hole.\footnote{Asymptotically Kaluza-Klein spacetimes containing black holes and bubbles have previously been considered in certain theories and similar contributions to the first law were found (see e.g. \cite{Kastor,Yz1,Yz2}).}

In this note we will revisit the derivation of the first law of black hole mechanics (and Smarr relation) in a general five dimensional theory of Einstein gravity coupled to an arbitrary number of Maxwell fields $F^I$ and neutral scalars $\chi^A$.  We assume the spacetime is asymptotically flat, stationary with two commuting rotational symmetries, and allow for a general $H_2(\Sigma)$ and the possibility of a black hole. Our main results are as follows.

 In the absence of a black hole we find that the mass of a soliton and the mass difference between nearby solitons satisfy
\bea
M &=& \frac{1}{2} \sum_{[C]} \Psi_I[C] q^I[C]  +E_\chi  \; ,\label{solmass} \\
\delta M &=& \sum_{[C]} \Psi_I[C] \delta q^I[C]  \; ,  \label{solmech}
\eea
where $[C]$ are a basis for $H_2(\Sigma)$,
\be\label{H_2:flux}
q^I[C] = \frac{1}{4\pi} \int_C F^I
\ee
are magnetic fluxes associated to each 2-cycle, $\Psi_I[C]$ are corresponding potentials which we define below, and $E_\chi$ is the potential energy of the scalars (this is of course absent in $U(1)^N$-supergravity).  Due to its formal similarity to the first law of thermodynamics,  we refer to (\ref{solmech}) as the {\it first law of soliton mechanics}. However, we emphasise that in contrast to the black hole case below, the analogy to a thermodynamic system is unclear. Curiously, the conserved angular momentum and electric charge do not appear explicitly in this law for solitons.  It is worth noting we also found an alternative form for these soliton mass formulas, see equations (\ref{BHmass}) and (\ref{BHmech}) specialised to the case with no black hole.

If a black hole is present in the spacetime we find that on top of the usual terms arising from the conserved charges of the spacetime, one also acquires extra terms associated to the 2-cycles and also certain discs that meet the horizon, see equations (\ref{BHmass}) and (\ref{BHmech}).  We note that for the standard black rings with no bubbles, this gives an alternative explanation for the origin of the dipole charge term, whereas for any more general black rings with exterior bubbles the dipole charge term is replaced by terms involving the magnetic flux through such discs.

We derive these results in section 2 and discuss them further in section 3.

\section{Mass and mass variation formulas}

\subsection{General theory}

We consider the general theory
\be\label{gentheory}
S =\tfrac{1}{16\pi} \int \star R - f_{AB}(\chi) \td\chi^A \wedge \star \td\chi^B- g_{IJ}(\chi) F^I \wedge \star F^J  - \star V(\chi) - \frac{1}{6}C_{IJK} F^I \wedge F^J \wedge A^K
\ee
where $F^I = \td A^I$ for a {\it locally} defined gauge potential $A^I$,  the couplings $f_{AB}, g_{IJ}$ are positive definite and $C_{IJK}=C_{(IJK)}$ are constants. This includes theories ranging from pure Einstein-Maxwell to the aforementioned supergravity theories. The Einstein equations and Maxwell equations are
\be
R_{ab} = g_{IJ} \left( F^I_{ac} F_{b}^{Jc} -\frac{1}{6} F^I \cdot F^J  g_{ab}\right)  +f_{AB} \partial_a \chi^A \partial_b \chi^B  + \frac{1}{3} V g_{ab}
\ee
and
\be
\td( g_{IJ} \star F^J) + \frac{1}{4}C_{IJK} F^J \wedge F^K  = 0  \; .
\ee
We will not need the explicit form of the scalar equation of motion.  We now turn to the derivation of the mass and mass variation formula.

The key idea in these derivations is to avoid the introduction of gauge potentials which, for non-trivial topology, are not globally defined. We follow the original approach by Bardeen, Carter and Hawking \cite{BCH}, hence our results are valid for perturbations which are invariant under the stationary and rotational Killing fields of the background.  It could be interesting to remove the symmetry restrictions on the perturbations using the elegant formalism of~\cite{Iyer:1994ys}. Although, we note that even with such a symmetry restriction, the space of solutions appears to be vast due to the possibility of arbitrary 2-cycle structure.  We will now sketch the derivation of our results. Many of the constructions and methods are well known, so we will be brief.

Consider an asymptotically flat spacetime $(M, g_{ab})$ with $\mathbb{R}\times U(1)^2$ symmetry and denote the corresponding stationary and rotational Killing fields by $K$ and $m_i$ for $i=1,2$ respectively.  If a black hole is present we assume the future horizon $\mathcal{H}^+$ is a Killing horizon with normal $\xi =K+ \Omega_i m_i$ with surface gravity $\kappa$ and angular velocities $\Omega_i$.  Let $\Sigma$ be a partial Cauchy surface, tangent to the rotational Killing fields $m_i$, which intersects spacelike infinity $S_\infty \cong S^3$. If there is no black hole present we assume $\Sigma$ is a Cauchy surface, whereas if there is a black hole we assume $\Sigma$ intersects the horizon $\Sigma \cap \mathcal{H}^+ =H$. Hence our arguments will be valid for both non-extremal and extremal black holes (we do not assume the existence of a bifurcation surface). We will treat the case with and without a black hole simultaneously, the only difference is in the former case the boundary $\partial \Sigma$ has an inner component given by $-H$. To recover the soliton case from the black hole case, one simply drops all horizon terms (and replaces $\xi$ with $K$).

The Komar mass and angular momenta are defined as usual
\be
M = -\frac{3}{32 \pi} \int_{S_\infty} \star \td K \; , \qquad \qquad J_i = \frac{1}{16\pi } \int_{S_\infty} \star \td m_i  \; .
\ee
By standard arguments, which invoke Stokes' theorem and the identity $\int_{H} \star \td\xi = -2 \kappa A_H$,  where $A_H$ is the area of $H$,  one can easily establish the so called Smarr relation
 \be\label{Smarr1}
M = \frac{3 \kappa A_H}{16 \pi} + \frac{3}{2} \Omega_i J_i  - \frac{3}{16 \pi} \int_\Sigma \star R(\xi)
\ee
where we have defined the Ricci 1-form $R(\xi)_a = R_{ab} \xi^b$.  

Let us now evaluate this Smarr relation for a solution $(g_{ab}, F^I_{ab}, \chi^A)$ of the general theory (\ref{gentheory}). Defining the 3-forms $G^I = \star F^I$ we may write the Einstein equations and Maxwell equations in the useful form
\be
R_{ab}= g_{IJ} \left( \frac{2}{3} F^I_{ac} F_{b}^{Jc} + \frac{1}{6} G^I_{acd} G_b^{Jcd}  \right)  +f_{AB} \partial_a \chi^A \partial_b \chi^B  +\frac{1}{3} V g_{ab}
\ee
and
\be
\nabla^b (g_{IJ} F^J_{ba}) + \frac{1}{8}C_{IJK} F^{J bc}G^K_{abc} = 0  \; .
\ee
Now suppose all the matter fields are invariant under the Killing field $\xi$. The condition $\mathcal{L}_\xi F^I=0$ implies
\be
i_\xi F^I = \td\Phi^I   \label{Phi}
\ee
for some globally defined ``electric" potentials $\Phi^I$ (recall by topological censorship $\Sigma$ must be simply connected). On the other hand, $\mathcal{L}_\xi G^I=0$ together with the Maxwell equations implies
\be
g_{IJ} i_\xi G^J = \Theta_I+ \frac{1}{2} C_{IJK} F^J \Phi^K   \label{Theta}
\ee
for globally defined 2-forms $\Theta_I$ which are closed but not necessarily exact (since we do not assume $H_2(\Sigma)$ is trivial).  Notice that $\Phi^I$ and $\Theta_I$ are invariant under the Killing field $\xi$. Using the field equations it follows that
\be
* R(\xi)  =  - \frac{1}{3} \Theta_I \wedge F^I  + \frac{2}{3} \td  (g_{IJ} \star F^I \Phi^J)  + \frac{1}{3}V \star \xi \; .
\ee
Hence we obtain the Smarr relation:
\be
M = \frac{3 \kappa A_H}{16 \pi} + \frac{3}{2} \Omega_i J_i + \frac{1}{16\pi} \int_\Sigma (\Theta_I \wedge F^I  - V(\chi)\star\xi )+ \frac{1}{8\pi} \int_H \Phi^I g_{IJ}  \star F^J  \label{smarr}
\ee
where we have fixed $\Phi^I$ to vanish at infinity, which by asymptotic flatness guarantees that $F^I, \Phi^I$ decay sufficiently fast so that the integral over $S_\infty$ makes no contribution. Note there are no terms from the scalar fields $\chi^A$ other than the bulk contribution from the potential $V(\chi)$, which we denote by $E_\chi \equiv-(1/16\pi)\int_\Sigma V \star \xi$. This reduces to the Smarr relation found for solitons in supergravity~\cite{Gibbons:2013tqa}.

Now consider a linearised perturbation $(\delta g_{ab}, \delta F^I_{ab}, \delta \chi^A)$  of the original solution which solves the linearised equations of motion and preserves the $\mathbb{R}\times U(1)^2$ symmetry. We write $h_{ab} = \delta g_{ab}$. To derive the mass variation law we follow the method originally developed in four dimensions, which employs the Smarr relation together with various less straightforward arguments~\cite{BCH, Heusler:1993cj, Heusler}. This begins by choosing a gauge so that the nearby solution has the same Killing fields so $\delta K^a=0$ and $\delta m_i^a=0$ and that the event horizon $\mathcal{H}^+$ is in the same position (this implies $\xi \wedge \delta \xi=0$ and $\mathcal{L}_\xi \delta \xi=0$). We also assume the non-trivial 2-cycles $ C \subset \Sigma$ are in the same position (we explain what the conditions for this are below).   Using the Smarr relation (\ref{Smarr1}) and the ADM mass formula for the perturbation
\begin{equation}
 \delta M = \frac{3}{8\pi}\int_{S_{\infty}} \star(K \wedge h)  \; ,
\end{equation}
where the 1-form $h_a = \tfrac{1}{2} ( \nabla^b h_{ba} - \nabla_a h^b_b)$, one can then prove a general version of the first law
\begin{equation}\label{massvariation1}
\delta M = \frac{\kappa \delta A}{8\pi} + \Omega_i  \delta J_i^H+\frac{1}{16\pi}\int_\Sigma G^{ab}h_{ab} \star K- \frac{1}{8 \pi}\delta \int_{\Sigma} \star G(K)  \; ,
\end{equation}
where the 1-form $G(K)_a=G_{ab}K^b$, the $G_{ab}$ is the Einstein tensor and $J^H_i$ is the Komar angular momentum defined on $H$.

We now turn to evaluating the mass variation formula \eqref{massvariation1} using the method developed in~\cite{Heusler:1993cj, Heusler}. After some manipulations involving the Einstein equation and scalar equation of motion, we find
\be
G^{ab} h_{ab} \star K - 2 \delta \star G(K)= 2g_{IJ} i_K G^I\wedge \delta F^J -2 i_KF^I \wedge \delta( g_{IJ} \star F^J)  -2 \td ( f_{AB} \delta \chi^A i_K\star d\chi^B)  \; .
\ee
It is clear the integral over $\Sigma$ of the scalar variation term can be converted to an integral over $\partial \Sigma$. Suppose we restrict to variations which do not change the asymptotic values of the scalars $\delta \chi^A_{\infty}=0$. Then, asymptotic flatness guarantees that $\delta \chi^A, \td\chi^A$ decay sufficiently fast so that the contribution from $S_\infty$ vanishes. Furthermore, it can be shown that the contribution from $H$ also vanishes. Hence we are left with
\bea \label{firstlaw:2}
\delta M &=& \frac{ \kappa \delta A_H}{8\pi} + \Omega_i  \delta J_i^H + \frac{1}{8\pi} \int_\Sigma g_{IJ} i_K G^I\wedge \delta F^J - i_K F^I \wedge \delta( g_{IJ} \star F^J) \nonumber \\ &=& \frac{ \kappa \delta A_H}{8\pi} + \Omega_i  \delta J_i + \frac{1}{8\pi} \int_\Sigma g_{IJ} i_\xi G^I\wedge \delta F^J - i_\xi F^I \wedge \delta( g_{IJ} \star F^J)   \; ,
\eea
where to obtain the second equality we have converted the Komar angular momentum over $H$ to the standard one over infinity $S_\infty$, used the non-trivial identity
\bea
&& \delta (R(m_i)) - i_{m_i} F^I \wedge \delta (g_{IJ} \star F^J) + g_{IJ} \delta F^I \wedge i_{m_i} \star F^J  \nonumber \\ &&\qquad = i_{m_i} \left( \frac{2}{3} g_{IJ} \delta F^I\wedge G^J - \frac{1}{3} F^I \wedge \delta (g_{IJ} G^J) +\frac{1}{3} \delta V \epsilon \right)   \; ,
\eea
and the fact that total $i_{m_i}$-derivatives vanish when integrated over $\Sigma$. Finally, using the Maxwell equation and the definitions (\ref{Phi}) and (\ref{Theta}) we obtain the first law
\be
\delta M = \frac{ \kappa \delta A_H}{8\pi} + \Omega_i \delta J_i + \frac{1}{8\pi} \int_\Sigma  \Theta_I \wedge \delta F^I  + \frac{1}{8\pi} \int_H \Phi^I \delta( g_{IJ} \star F^J)   \label{1stlaw}
\ee
where we have again assumed $\Phi^I$ vanishes at infinity. We emphasise that there are no terms arising from the variation of the scalar fields and in contrast to the Smarr relation the scalar potential has cancelled (although had we allowed variations in the moduli $\chi^A_\infty$ there would be such terms).

It is worth noting that the Smarr relation (\ref{smarr}) and first law (\ref{1stlaw}) we have so far obtained  make use of the assumption of $U(1)^2$-rotational symmetry only if a black hole is present. Hence the mass (\ref{smarr}) and mass variation law (\ref{1stlaw}) for solitons, which reduce to a single integral over $\Sigma$, remain valid for general stationary spacetimes (with $\xi$ replaced by $K$). In fact we may evaluate these bulk integrals more explicitly by making more detailed use of the $U(1)^2$-rotational symmetry.

\subsection{Exploiting rotational symmetry}

We may use various well-known methods for (generalised) Weyl solutions with Maxwell fields, see e.g.~\cite{Kunduri:2006uh, Tomizawa:2009ua}. Given any two commuting Killing fields $X,Y$ which leave the Maxwell fields invariant, one can show using the Bianchi identities that the functions  $i_Xi_Y F^I$ are constants. Asymptotic flatness implies that a different linear combination of the $m_i$ vanishes on the two axes extending to infinity. Hence we deduce that $i_\xi i_{m_i} F^I=0$ and $i_{m_1}i_{m_2} F^I=0$. Similarly, invariance of the fields together with the Maxwell equations implies $i_\xi i_{m_1} i_{m_2} g_{IJ} G^J$ are constant functions which must therefore also vanish. Furthermore, $\mathcal{L}_{m_i} F^I$=0 also allows one to establish the existence of globally defined ``magnetic" potentials $\Phi^I_i$ such that
\be\label{spatialpot}
i_{m_i} F^I = \td \Phi^I_i  \; .
\ee
We deduce that the magnetic potentials $\Phi^I_i$ and electric potentials $\Phi^I$ are invariant under all the Killing fields $\xi, m_1, m_2$. From (\ref{Theta}) it follows that $\mathcal{L}_{m_i}\Theta_I=0$ and since by definition $\td\Theta_I=0$, it is easy to see that $\td (i_{m_i} \Theta_I)=0$.  Hence there exists a globally defined potential $U_{Ii}$ such that
\be \label{dU}
i_{m_i} \Theta_I= \td U_{Ii} -  \frac{1}{2} C_{IJK} \td\Phi_i^J \Phi^K_{H} \; ,
\ee
where we have chosen to add the total derivative on the RHS for later convenience and $\Phi^I_H$ are the electric potentials evaluated on the horizon (we show below these are constants).  On the other hand, contracting (\ref{Theta}) gives
\be
i_{m_i} \Theta_I = g_{IJ} i_{m_i} i_\xi G^J - \frac{1}{2} C_{IJK} \td\Phi_i^J \Phi^K  \; .
\ee
It follows that the potentials $U_{Ii}$ are also invariant under all the Killing fields $\xi, m_1,m_2$.

For later reference, consider the behaviour of the potentials at an axis or at a horizon. It is clear that along an axis where say $m_i=0$  the corresponding potentials $\Phi^I_i, U_{Ii}$ must be constant. Next consider the potentials on the horizon $\mathcal{H}^+$. Since in general $\text{Ric}(\xi,\xi)|_{\mathcal{H}^+}=0$ we deduce that $i_\xi F^I$ is null on the horizon, and since it is also tangent to the horizon (i.e. it is orthogonal to $\xi$), we deduce that $i_\xi F^I |_{\mathcal{H}^+}\propto \xi$. This shows that the electric potentials $\Phi^I$ are constant on the horizon. Similarly, $\text{Ric}(\xi,\xi)|_{\mathcal{H}^+}=0$ also implies the 2-form $i_\xi G^I$ must be null and tangent on the horizon. This means that $i_\xi  G^I|_{\mathcal{H}^+} \propto \xi \wedge X$ where $X$ is tangent to a cross-section $H$. Hence $i_{m_i} i_\xi  G^I|_{\mathcal{H^+}} \propto \xi$ and so we deduce that $\td U_{Ii}|_{\mathcal{H}^+} \propto \xi$ and hence the potentials $U_{Ii}$ are all constant on the horizon.

The above considerations allow us to express the Maxwell fields $F^I$ and their duals $G^I$ in terms of the potentials $\Phi^I, \Phi_i^I, U_{Ii}$. Due to the invariance of these potentials under the $\mathbb{R}\times U(1)^2$ isometry, they can be thought of as functions on the 2d orbit space $B \cong \Sigma / U(1)^2$. The orbit space $B$ is a simply connected manifold with boundary $\partial B$ and corners, see e.g.~\cite{Hollands:2007aj, Hollands:2007qf}. In the interior, on the 1d boundary segments (except the part corresponding to $H$) and at the corners, the matrix of Killing fields $m_i \cdot m_j$ has rank $2,1,0$ respectively. We note the orbit space $B$ is non-compact with one asymptotic end corresponding to spatial infinity.\footnote{This is also true for the extremal case since we have chosen $\Sigma$ to intersect the future horizon. In the extremal case one could choose $\Sigma$ to not intersect the horizon, in which case it has an asymptotically cylindrical end which would be inherited by the orbit space~\cite{Figueras:2009ci}.}  The boundary $\partial B \cong \mathbb{R}$ is divided into intervals $I$, or ``rods", all of which are compact except for the two asymptotic ends. If a black hole is present, the horizon, which we assume to be connected, corresponds to a compact interval $I_H \cong H/U(1)^2$. The rest of  $\partial B$ is divided into intervals which correspond to surfaces where some integer linear combination of the Killing fields $m_1$ and $m_2$ vanish. The two non-compact ``semi-infinite" intervals correspond to the axes of rotation in the spacetime which extend out to spatial infinity. 

Now, recall that for the perturbed solution we have assumed the Killing fields and the position of the horizon and 2-cycles are unchanged. Therefore, it follows that the orbit space of the perturbed solution is the same as that of the background solution (including its boundary segments and corners).  Hence we may reduce both  integrals \eqref{smarr} and \eqref{1stlaw} to integrals of the potentials $\Phi^I, \Phi^I_i, U_I$ and their perturbations over $B$. This can be most efficiently done by noting that for any 4-form $\omega$ on $\Sigma$ which is invariant under the Killing fields $m_i$, we have $\int_{\Sigma} \omega  = 2\pi^2 \int_{B} \eta_{ij} i_{m_j} i_{m_i} \omega$, where $\eta_{ij}$ is the antisymmetric symbol with $\eta_{12}=1$ and we have chosen the $m_i$ to have $2\pi$-periodic orbits.  

A short calculation reveals that the terms in the Smarr relation (\ref{smarr}) reduce to
\begin{equation}\label{smarr:pot}
\frac{1}{16\pi} \int_\Sigma \Theta_I \wedge F^I + \frac{1}{8\pi} \int_H \Phi^I g_{IJ}  \star F^J = \Phi_H^I Q_I + \frac{\pi}{4}\int_{B} \eta_{ij} \td U_{Ij} \wedge \td\Phi^I_i  \; ,
\end{equation} 
where we have used the Maxwell equations, defined the electric charges
\begin{equation}
Q_{I} \equiv \frac{1}{8\pi}\int_{S_{\infty}} g_{IJ}\star F^J \; ,
\end{equation} 
and used equation (\ref{dU}). A similar calculation reveals that the terms in the first law (\ref{1stlaw}) reduce to
\begin{equation} \label{1stlaw:pot}
\frac{1}{8\pi} \int_\Sigma  \Theta_I \wedge \delta F^I  + \frac{1}{8\pi} \int_H \Phi^I \delta( g_{IJ} \star F^J) = \Phi_H^I \delta Q_I + \frac{\pi}{2}\int_{B} \eta_{ij}\td U_{Ij} \wedge \td\delta \Phi^I_i    \; .
\end{equation}
Observe that the integrands of the integrals over $B$ are total derivatives and hence they further reduce to integrals over its  boundary $\partial B$ and its asymptotic end. By asymptotic flatness of the spacetime we impose that all the potentials vanish at infinity, in which case we deduce that the contribution from the asymptotic end vanishes. Hence we may write these integral in two ways,
\be
\int_B  \eta_{ij} \td U_{Ij} \wedge \td\Phi^I_i  =  \int_{\partial B} \eta_{ij} U_{Ij} \td\Phi^I_i  = -  \int_{\partial B} \eta_{ij} \Phi^I_i \td U_{Ij}    \label{dB}
\ee
and similarly for the analogous perturbation term. We will use both of these forms below.  We now evaluate these integrals over each boundary segment.

Without loss of generality we may choose the two semi-infinite rods to correspond to the vanishing of $m_1$ and $m_2$ respectively. From \eqref{spatialpot} and \eqref{dU} we deduce that the corresponding  potentials $( \Phi_i^I, U_{Ii})$ must vanish all the way along the associated semi-infinite rod (since they are constant and vanish at infinity). Using either expression (\ref{dB}), it follows that the contributions to the integrals \eqref{smarr:pot} and \eqref{1stlaw:pot} from the two semi-infinite axes must vanish.

Now consider a compact rod $I$ defined by $v_i m_i=0$ for some $v_i \in \mathbb{Z}$, which is not adjacent to $I_H$. We introduce a new basis $\hat{m}_i = S_{ij} m_j$ where $S_{1i}=v_i$ and let $S_{2i}=w_i$, so $\hat{m}_1 = v_i m_i$ and $\hat{m}_2  = w_i m_i$. We wish to preserve the $2\pi$ periods of these Killing fields and hence require $S \in SL(2,\mathbb{Z})$.  The rod $I$ corresponds to the surface defined by $\hat{m}_1=0$ on which $\hat{m}_2$ is non-vanishing everywhere except at the two points corresponding to the endpoints of $I$. Hence such rods correspond to topologically $S^2$ submanifolds in the spacetime, i.e. they are  2-cycles $C \subset \Sigma$ which correspond to non-trivial elements  $[C] \in H_2(\Sigma)$.  Hence
\be
\int_{I} \eta_{ij} U_{Ij} \td \Phi^I_i  = \int_{I} \eta_{ij} \hat{U}_{Ij} \td \hat{\Phi}^I_i = - \hat{U}_{I 1} \int_I \td \hat{\Phi}^I_2 = \frac{2}{\pi} \Psi_{I }[C] q^I[C]
\ee
where the first equality follows from $SL(2,\mathbb{Z})$ invariance, in the second equality we have used the fact that the potentials $\hat{\Phi}^I_1, \hat{U}_{I 1}$ are constants on $I$, and in the final equality we have defined the potential $\Psi_I[C] \equiv \pi \hat{U}_{I1}$ and used the definition \eqref{H_2:flux} of the fluxes $q^I[C]$. Observe that in terms of the integers $v_i$ which specify which linear combination of $m_i$ vanishes on $C$ we have
\be
\Psi_I[C] = \pi v_i U_{Ii}  \; .
\ee
Similarly we find
\be
\int_{I} \eta_{ij} U_{Ij} \td\delta \Phi^I_i  = \frac{2}{\pi} \Psi_I[C] \delta q^I[C]  \; .
\ee 
Therefore, if the spacetime does not contain a black hole, equations \eqref{smarr} and \eqref{1stlaw} give the soliton mass formula \eqref{solmass} and first law of soliton mechanics \eqref{solmech}.

Now suppose there is a black hole in the spacetime. In this case it appears more natural to evaluate the second expression in (\ref{dB}). Indeed, since $U_{Ii}$ are constants on the horizon, there is no contribution from the part of the integral along $I_H$. Consider a compact rod $I_D$ defined by $\hat{m}_1 = v_i m_i=0$, which is adjacent to the horizon rod $I_H$.  As in the previous case, there is a Killing field $\hat{m_2}$ non-vanishing on the surface corresponding to the interior of $I_D$, however it vanishes only at one point corresponding to the endpoint of $I_D$ not shared by $I_H$. Hence $I_D$ corresponds to a disc topology surface $D$ in the spacetime. Note there can be at most two such surfaces corresponding to compact rods on either side of $I_H$.  Now we define a magnetic flux associated to such a disc by
 \be
 { \cal Q}_I[D] = \frac{1}{4} \int_{D} \left( \Theta_I+ \frac{1}{2}C_{IJK} F^J \Phi^K_H \right)  \; .   \label{flux}
 \ee
 Observe that this flux can be evaluated over any representative of $[D]$, i.e. surfaces which are homologous to $D$ with the same boundary as $D$.
 Then we have
 \be
 \int_{I_D} \eta_{ij} \Phi^I_i \td U_{I j} = \hat{\Phi}^I_1 \int_{I_D} \td \hat{U}_{I 2} =- \frac{2}{\pi} \Phi^I[D] {\cal Q}_I[D]
 \ee
 where the first equality follows from the fact that $\hat{\Phi}^I_1, \hat{U}_{I1}$ are constants on $I_D$, and in the second we have defined $\Phi^I[D] \equiv \hat{\Phi}_1^I$ and used (\ref{dU}) to evaluate (\ref{flux}). Observe that in terms of the integers $v_i$ which define the disc
 \be
 \Phi^I[D] \equiv  v_i \Phi^I_i  \; .
 \ee
 Similarly, one finds
 \be
 \int_{I_D} \eta_{ij} \delta \Phi^I_i \td U_{I j} = -\frac{2}{\pi} \delta \Phi^I[D] {\cal Q}_I[D]   \; .
 \ee
 The compact rods which are not adjacent to $I_H$ correspond to 2-cycles $C$ as above and in this case one can evaluate the integrals in the same way as for the discs, with analogous definitions of $\Phi^I[C]$ and ${\cal Q}_I[C]$. Putting everything together, we see that  equations \eqref{smarr} and \eqref{1stlaw} give the following black hole mass formula
\be
 M = \frac{3 \kappa A_H}{16 \pi} + \frac{3}{2} {\Omega}_i  {J}_i + \Phi^I_HQ_I+  \frac{1}{2}\sum_{[C]}  {\cal Q}_I[C]  \Phi^I[C]+   \frac{1}{2}\sum_{[D]}{\cal Q}_I[D]   \Phi^I[D]  +E_\chi \label{BHmass}
 \ee
 and finally the first law of black hole mechanics
 \be
 \delta M = \frac{ \kappa \delta A_H}{8\pi} + \Omega_i \delta J_i + \Phi^I_H \delta Q_I + \sum_{[C]} {\cal Q}_I[C]  \delta  \Phi^I[C] +\sum_{[D]}  {\cal Q}_I[D]  \delta \Phi^I[D]    \; .  \label{BHmech}
\ee
Observe that in the absence of a black hole, in which case there are also no disc terms, this gives an alternate form for the mass and first law for solitons (\ref{solmass}) and (\ref{solmech}).  We will discuss how this general first law reduces to the various known cases below.

\section{Discussion}

We have seen that additional terms arise in the Smarr relation (\ref{BHmass}) and first law of black hole mechanics (\ref{BHmech}) which are solely due to the Maxwell fields coupled to the non-trivial topology of the spacetime.  From our derivation, it is clear that the presence of Chern-Simons (CS) terms is not necessary for this phenomenon, and indeed, contributions from the 2-cycles are still present in pure Einstein-Maxwell theory (although we are not aware of an explicit solution with bubbles in this case). For BPS solitons in supergravity it was found that the CS terms appear to be (partially) responsible for their existence~\cite{Gibbons:2013tqa}, however, since the CS terms are necessary for {\it any} (non-static) BPS solution it is not really possible to deduce this solely from their study. By working with non-BPS solutions, or even solutions to theories with no supersymmetry, we have found the basic reason such solitons exist is non-trivial spacetime topology. Thus we see no reason why such solitons cannot exist in theories with no CS terms. It would be interesting to construct such solitons in pure Einstein-Maxwell theory to confirm this explicitly.

Curiously, the form of the first law for black holes (\ref{BHmech}) exchanges the usual role of a charge $\mathcal{Q}_I$ and a potential $\Phi^I$. One can remedy this by performing a Legendre transform $M' = M -\sum_{[C]} {\cal Q}_I[C]   \Phi^I[C]-  \sum_{[D]} {\cal Q}_I[D]  \Phi^I[D]$ so that $\delta M'$ is as (\ref{BHmech}) except the 2-cycle and disc terms become of the standard form $-\Phi^I \delta \mathcal{Q}_I$. It would be interesting if $M'$ has some physical significance.

We now discuss how our general Smarr relation (\ref{BHmass}) and first law (\ref{BHmech}) reduce to the known special cases which have been previously studied. Most simply, consider the horizon topology $H \cong S^3$ with a trivial $H_2(\Sigma)=0$ so that $\Sigma \cong \mathbb{R}^4 -B_4$ where $B_4$ is a 4-ball. Then there are no 2-cycles $C$ or discs $D$ and we lose the last two sets of terms, recovering the Smarr relation and first law presented in~\cite{Gauntlett:1998fz}.

On the other hand, for a black ring $H \cong S^1 \times S^2$ with no bubbles outside, there exists one disc $D$ and one can show $\Sigma \cong \mathbb{R}^4 - (S^1 \times B_3) \cong D \times S^2-  \{pt\}$~\cite{Alaee:2013oja}. Then we see that an extra term arises in the Smarr relation and first law, which corresponds to the known dipole charge term~\cite{Emparan:2004wy, Copsey:2005se}, as we now show.  Without loss of generality, suppose $m_1=0$ on $D$. Then using the various constancy properties of the potentials, which together with our asymptotic conditions imply the potentials $\Phi_i^I, U_{I i}$ vanish on the axis extending out to infinity along which $m_i=0$, it is easy to see that  ${\cal Q}_I[D] = \pm \tfrac{1}{2}\pi U_{I2}|_H$ and  $\Phi^I[D] =\mp 2 q^I$, where $q^I = \tfrac{1}{4\pi} \int_{S^2} F^I$ is the dipole charge defined on the $S^2 \subset H$. Hence, defining the horizon magnetic potential $\Psi_I^H \equiv  -\pi U_{I 2}|_H$, we see that the terms from $D$ in the Smarr relation (\ref{BHmass}) and first law (\ref{BHmech}) are $\tfrac{1}{2}\Psi^H_I q^I$ and $\Psi^H_I \delta q^I$ respectively, as required. We note that this gives a more straightfoward derivation of the first law for black rings (with no bubbles) revealing that the origin of the dipole term naturally arises from the flux through a disc ending on the horizon. Interestingly, for black rings with external bubbles it appears that one cannot express the contributions from the discs solely in terms of the dipole charge defined on the $S^2$ of the horizon.

In general, with the assumed symmetries the horizon topology $H$ must be one of  $S^3, S^1\times S^2, L(p,q)$~\cite{Hollands:2007qf}. Furthermore, for a general $H_2(\Sigma)$,  the topology of the domain of outer communication is $\Sigma \cong ( \mathbb{R}^4 \# n  (S^2\times S^2)  \# n' (\pm \mathbb{CP}^{2})) - {\cal B}$ where $\partial {\cal B} = H$~\cite{Hollands:2010qy}. Our mass and first law possess extra terms for any black hole horizon topology, even spherical topology. The terms coming from the 2-cycles $C$ may be interpreted as corresponding to the mass of the bubbles, whereas the terms coming from the discs $D$ may be interpreted as an interaction term between a bubble and the black hole. 

It is clear that black holes in an asymptotically flat background spacetime containing bubbles represent a gross violation to black hole uniqueness, on top of the existence of black rings. Indeed, this applies even to spherical topology black holes.\footnote{The black hole uniqueness theorem~\cite{Reall:2002bh} assumes the stationary Killing field is timelike everywhere outside the horizon. However, for the known solitons the stationary Killing field is not strictly timelike everywhere and hence one may evade this theorem.}  Presumably, as for vacuum gravity~\cite{Hollands:2007aj, Figueras:2009ci}, for theories with hidden symmetry (such as supergravity) one can demonstrate a uniqueness theorem for a given 2-cycle structure by specifying additional data such as the various magnetic fluxes. Indeed, there are partial results in Einstein-Maxwell theory~\cite{Hollands:2007qf}, although for minimal supergravity the results so far have been restricted to trivial spacetime topology~\cite{Tomizawa:2009ua, Tomizawa:2009tb} or multiple horizons~\cite{Armas:2009dd}. Unlike the vacuum case though, it is easier to envisage the existence of a vast set of black hole solutions with non-trivial bubbles supported by magnetic flux.  Our first law provides a simple starting point for studying the space of such black holes.

Finally, it would be interesting to explore the microscopic significance of these extra contributions to the first law of black hole thermodynamics.   \\

\noindent {\bf Acknowlegments}: We would like to thank Centro de Ciencias de Benasque Pedro Pascual for hospitality, where this work was initiated. We would also like to thank Joan Simon and Nicholas Warner for useful discussions. HKK is supported by an NSERC Discovery Grant. JL is supported by an EPSRC Career Acceleration Fellowship.

\end{document}